\documentclass[12pt]{elsart}
\usepackage{epsfig}
\journal{ }

\def\epj#1#2#3#4{    {Eur. Phys. J. }#1 {\bf #2} (#3) #4}
\def\np#1#2#3#4{    {Nucl. Phys. }#1 {\bf #2} (#3) #4}
\def\npp#1#2#3{    {Nucl. Phys. Proc. Suppl. }{\bf #1} (#2) #3}
\def\pl#1#2#3#4{    {Phys. Lett. }#1 {\bf #2} (#3) #4}

\def\pr#1#2#3#4{    {Phys. Rev. }#1 {\bf #2} (#3) #4}

\def\prl#1#2#3{    {Phys. Rev. Lett. }{\bf #1} (#2) #3}

\def\mpl#1#2#3#4{   {Mod. Phys. Lett. }#1 {\bf #2} (#3) #4}

\def\re#1{{\mathrm Re}\left\{#1\right\} }
\def\im#1{{\mathrm Im}\left\{#1\right\} }
\newcommand{\com}[1]{ \par }
\def\spa{\hspace*{-.65cm}}

\def\evg{\, g_{eV}^\gamma}
\def\tvg{\, g_{\tau V}^\gamma}
\def\evz{\, g_{eV}^Z}
\def\eaz{\, g_{eA}^Z}
\def\tvz{\, g_{\tau V}^Z}
\def\taz{\, g_{\tau A}^Z}
\def\tz{\, g_{\tau T}^Z}
\def\tg{\, g_{\tau T}^\gamma}
\def\evg2{(g_{eV}^\gamma)^2}
\def\tvg2{(g_{\tau V}^\gamma)^2}
\def\evz2{(g_{eV}^Z)^2}
\def\eaz2{(g_{eA}^Z)^2}
\def\tvz2{(g_{\tau V}^Z)^2}
\def\taz2{(g_{\tau A}^Z)^2}
\def\tz2{(g_{\tau T}^Z)^2}
\def\tg2{(g_{\tau T}^\gamma)^2}

\def\re#1{{\mathrm Re}\left\{#1\right\} }
\def\im#1{{\mathrm Im}\left\{#1\right\} }

\def\dps{\displaystyle}
\def\spa{\hspace*{-.65cm}}
\newcommand{\beq}{\begin{equation}}
\newcommand{\eeq}{\end{equation}}
\newcommand{\bi}{\begin{itemize}}
\newcommand{\ei}{\end{itemize}}
\newcommand{\bea}{\begin{eqnarray}}
\newcommand{\eea}{\end{eqnarray}}
\newcommand{\bes}{\begin{eqnarray*}}
\newcommand{\ees}{\end{eqnarray*}}

\begin{document}
\begin{flushright}FTUV-07-1607\end{flushright}
\begin{frontmatter}
\title{Tau anomalous magnetic moment form factor\\
at Super B/Flavor factories.}
\author[Valencia]{J. Bernab\'eu},
\author[Montevideo]{G. A. Gonz\'alez-Sprinberg}
\author[Valencia]{J. Papavassiliou}
\author[Valencia]{J. Vidal}
\address[Valencia]{Departament de F\'{\i}sica Te\`orica
Universitat de Val\`encia, E-46100 Burjassot,Val\`encia, Spain\\
and\\
IFIC, Centre Mixt Universitat de Val\`encia-CSIC, Val\`encia, Spain}
\address[Montevideo]{Instituto de F\'{\i}sica,
 Facultad de Ciencias, Universidad de la Rep\'ublica,
 Igu\'a 4225, 11400 Montevideo, Uruguay}

\journal{Nuclear Physics B}
  
\begin{abstract}

The proposed high-luminosity B/Flavor factories offer new opportunities
for the improved  determination of the fundamental physical parameters
of  standard heavy
leptons.  Compared to the electron  or the muon case,
the magnetic properties  of the $\tau$ lepton
are largely unexplored.  We show that
the electromagnetic  properties  of the $\tau$, and in particular
its magnetic form factor,
may be measured competitively
in these facilities, using unpolarized  or polarized  electron  beams.
Various observables of the $\tau$'s  produced on top of
the $\Upsilon$ resonances,
such as cross-section and normal polarization
for unpolarized electrons or
longitudinal and transverse asymmetries for
polarized beams, can be combined in order to increase the sensitivity on the
magnetic moment form factor. In the case of polarized electrons, we identify a special 
combination of transverse and longitudinal $\tau$ polarizations able to disentangle 
this anomalous magnetic  
form factor from both the charge form factor and the interference  
with the Z-mediating amplitude. For an  integrated  luminosity of $15 \times 10^{18} b^{-1}$ one could 
achieve a sensitivity of about $10^{-6}$, which is several orders  
of magnitude below any other
existing high- or low-energy bound 
on the magnetic moment. Thus one may obtain  
a QED test of this fundamental quantity to a few \% precision.
\end{abstract}
\end{frontmatter}

\section{Introduction}
Magnetic  properties  of  elementary  particles are  of  pa\-ra\-mount
importance   both  to   theoretical  and   experimental   high  energy
physics. The electron anomalous magnetic moment \cite{pdg} is measured
with the highest  precision available in physics nowadays:
\beq a_e =
\mu_e/\mu_B - 1 = \frac{g_e-2}{2} =
=  (1159.652 \, 181 \,0 \pm 0.000\, 000 \,7 )\times 10^{-6}
\eeq
and the best determination of  the fine structure
constant  $\alpha$  is  obtained  from this  electron  property.   The
theoretical  predictions  in   Quantum  Electrodynamics  (QED),
obtained by Schwinger in  1948 \cite{sch}, was the first one-loop
computation in quantum field theories to  be confronted
with experiment:
\beq a_e
= \frac{\alpha}{2 \pi} = \simeq 0.00116 \,.
\label{schw}\eeq
Actually, precision measurements of the anomalous magnetic moments
do not only test QED (up to 4-loops),
 but the electroweak  and hadronic interactions as well.
 In fact, since anomalous magnetic moments are chirality flipping  quantities,
weak quantum corrections to the
 muon  anomalous magnetic moment are enhanced
by a factor  of the  order of
 $m_\mu^2/m_e^2 \simeq 40000$ compared to that of the electron;
the measured value is~\cite{mue}
\bea a_\mu &=& \mu_\mu/(e\hbar/2 m_\mu) -
 1 = \frac{g_\mu-2}{2} \nonumber\\
&=& ( 1165.920 \,80 \pm 0.000\, 54 \pm 0.000\,33 ) \times 10^{-6}\,.
 \eea
In addition, new  physics models  \cite{kin}, and especially
those furnishing mass-enhancements comparable
to that of the Standard Model (SM),
can be constrained from  these extremely
 precise  measurements,
nowadays a subject of intense activity \cite{mut}.

In comparison with these values, our experimental knowledge on the magnetic moment of the $\tau$ lepton
is rather poor.
While the electron and muon anomalous magnetic moments are known with more than seven
figures, the PDG limit for the $\tau$ lepton magnetic moment anomaly is \cite{pdg} :
\beq
 - 0.052 \,< \, a_\tau \,  <\, 0.013 \,\,\;(95\,\%\, C.L.)
\label{mu}\eeq
These numbers are  more than one order of magnitude bigger than  the first-order
QED contribution, given in Eq.~(\ref{schw}). Recent computations have
re-analyzed both weak and hadronic corrections for the $\tau$ lepton magnetic anom\-aly within the
SM, with an excellent agreement with previous computations~\cite{passera}. The
higher-order QED contributions are at the level of 1\%
compared to  the one-loop result of Schwinger,
while   hadronic and weak corrections are at the level of 0.001\% and 0.04\%, respectively.

The experimental determination of the anomalous magnetic moment of the
fast-decaying $\tau$  is very different from that of the stable or
relatively  long-lived electron  and muon, simply because one does not have the
time to  measure  its interaction with  an
external  electromagnetic  field.
Instead, the  magnetic information
is carried by the cross-section or partial widths for the
$\tau$ pair production, together with
spin  matrices or angular distributions of the $\tau$ decay products.
For example, the PDG bounds in  Eq.(\ref{mu}) where obtained by the DELPHI
Collaboration~\cite{exp} from  the LEP2 data for
the total cross-section for the reaction
$e^+ e^- \rightarrow e^+  e^- \tau^+  \tau^-$, assuming that any deviation
from the tree level SM prediction was exclusively due 
to magnetic anomaly~\cite{cornet}.

The magnetic  anomaly, together  with analogous quantities  related to
the weak magnetism, {\it i.e.}  the magnetic coupling with the Z, have
been  already   investigated  experimentally~\cite{exp,heister}.   The
contributions to the magnetic and weak magnetic anomalies from physics
beyond the SM  have also been studied in the context  of both low- and
high-energy  physics~\cite{arcadi}.  It is  important to  emphasize at
this  point that, strictly  speaking, the  magnetic moment  anomaly is
defined with all three fields  entering into the interaction vertex on
their   mass-shell.   However,  in   several  of   the  aforementioned
experiments the  kinematics are such  that the $\tau$'s or  the photon
are in fact {\it off-shell};  therefore, what one actually measures is
the  corresponding  form-factor  (for   some  value  of  the  momentum
transfer) rather than $a_{\tau}$ itself.  This is usually accomplished
under  the additional assumption  that, of  all possible  form factors
appearing  in  the off-shell  vertex,  the  one  corresponding to  the
magnetic  moment gives  the dominant  contribution.   Such experiments
furnish {\it bounds} on the contributions to the magnetic moments from
physics beyond the SM (since the  scale of new physics is rather high,
these latter contributions are practically ``on-shell'').  This is the
point  of  view adopted  in~\cite{arcadi},  where  the most  stringent
model-independent limit for the  magnetic properties is obtained: \beq
- 0.007  \,  <  \,  a_\tau^{{\rm   New  Phys.}}   \,  <  \,  0.005\,\,
(\,95\,\,\%\,\, C.L.)  \eeq

The one-loop SM contribution to  the magnetic form factor was computed
in  ~\cite{fuji} long  time ago,  for arbitrary  values of  the photon
``off-shellness'' ($q^2$) and with the charged fermions on-shell.  The
form   factor  obtained  depends   on  the   electroweak  gauge-fixing
parameter~\cite{PT}, and  becomes gauge-independent only  in the limit
$q^2  \to 0$.   On the  other hand,  the pure  QED corrections  to the
off-shell form  factor are {\it  automatically} gauge-independent, for
any value of $q^2$.

 LEP has been the main  source of data on $\tau$-pair production until
the  advent of  the  B factories  and  their upgrades.
In the near  future they are
expected to  be superseded  by several  orders of
magnitude,  thanks to  the  high-luminosity Super  B Factories,  where
$10^{11}\sim  10^{12}$ $\tau$  pairs  will be  produced~\cite{superb}.
Motivated  by  these  possibilities,  CP-odd  spin  correlations  have
already   been  studied   in~\cite{Bernabeu:2004ww},   for  low-energy
physics.     In   addition,   polarized    beams   are    also   being
considered~\cite{superb};   their  implementation   would   allow  the
possibility of defining and  measuring new observables, related to the
linear $\tau$  polarization, not yet considered in  the literature with
respect to the magnetic properties.

In  this paper  we propose  new observables  in order  to  explore the
poorly  known magnetic  properties  of the  $\tau$  lepton, using  the
highest  statistics  facilities available  nowadays  and  in the  near
future.   Cross-sections  and   asymmetries   for  resonant   $\tau$
pair-production on  top of the $\Upsilon$ resonances  are studied.  In
particular  we show  that some  of  these observables,  with the  same
discrete  symmetry transformations as  the magnetic moment,
  allow one  to measure
these  properties with  a precision  up to  $10^{-6}$. We  compute the
contribution of the magnetic form  factor to the cross-section and the
normal  linear $\tau$ polarization  --for unpolarized  electron beams--
and both transverse and longitudinal $\tau$ polarizations for polarized
$e^{-}$-beams.

The fact that  in this class of experiments one  will be measuring the
magnetic  form-factor  rather  than  the  anomaly  provides  a  unique
opportunity to observe strong flavor-dependent effects, encoded in the
momentum-dependence of the  form factor, in the context  of {\it pure}
QED.   Indeed, whereas  the  Schwinger correction  (the leading  order
value  of   the  magnetic  form   factor  at  $q^2=0$)   is  universal
(i.e.  independent of  the fermion  masses), the  running of  the form
factors depends strongly  on the mass of the  fermion interacting with
the  photon.  Thus,  whereas  the corresponding  electron form  factor
practically  vanishes  at  the   values  of  $q^2$  that  we  consider
($\Upsilon$ resonance), the $\tau$ form factor drops only to about one
quarter of its initial value, because the heavy $\tau$ mass slows down
the running considerably. In  addition, for $q^2>4m^2_{\tau}$ the form
factor develops an imaginary part,  which, as we will demonstrate, can
also be experimentally measured.

The paper  is organized as follows:  In section (2)  the magnetic form
factors  are defined,  and  their one-loop  QED  prediction (real  and
imaginary parts) reported.  In  section (3) the $\tau$ pair production
at Super B factories is  studied, and expressions for the differential
cross-section and  the normal asymmetry  are derived.  In  section (4)
polarized beams  observables are considered,  with particular emphasis
on the  transverse and longitudinal  asymmetries.  In section  (5) the
advantages of  operating at  the $\Upsilon$ resonances  are discussed,
and the  role of  the weak contributions,  especially that of  the $Z$
boson,  considered.  Finally, in  section (6) we estimate the 
sensitivity expected for the anomalous magnetic moment form factor 
obtained from these observables, and present  our
conclusions and final remarks.

\section{$f\bar{f}\gamma$ vertex form factors}

The most general Lorentz invariant structure describing the
interaction of a vector boson $V$  with two on-shell fermions $f\bar{f}$ can be
written in terms of six form factors:
\begin{eqnarray}
&&\langle f(p_-)\bar{f}(p_+)|\,J^\mu(0)\,|0 \rangle=  e\; \bar{u}(p_-) \Big[ (F_1+ F_4 \gamma_5)\gamma^\mu\nonumber\\
&&\qquad\qquad\qquad+\frac{1}{2m_f}(i\, F_2+F_3\gamma_5)\sigma^{\mu\nu}q_\nu
+\frac{1}{2m_f}(i\, F_5+F_6\gamma_5)q^\mu\ \Big]v(p_+)
\label{eq:1}
\end{eqnarray}
where $q=p_++p_-$. Since the two fermions are on-shell
the form factors $F_i$ appearing in Eq.~(\ref{eq:1})
are functions of $q^2$ and $m_f^2$ only.

In addition,  if the current $J^\mu$ is conserved, we must have
\beq
i\frac{q^2}{2m_f}\, F_5+\left(\frac{q^2}{2m_f}F_6-2m_f \,F_4\right)\gamma_5=0\quad \Rightarrow\left\{\begin{array}{ll}
F_5=0\\
F_6=\dps\frac{4m_f^2}{q^2}F_4\end{array}\right.
\eeq
so that the final expression for the gauge invariant $f\bar{f}\gamma$ vertex reduces to:
\bea
&&\spa \langle f(p_-)\bar{f}(p_+)|\,J^\mu(0)\,|0\rangle= \nonumber\\
&& e\, \bar{u}(p_-) \left[  \gamma^\mu\; F_1
+\frac{1}{2m_f}(i\, F_2+F_3\gamma_5)\sigma^{\mu\nu}q_\nu +\left(q^2\gamma^\mu-q^\mu\not{\! q}\right)\gamma_5 F_A\right]v(p_+)
\label{eq:2}
\eea

In this expression, 
$F_1$ parametrizes the vectorial part of the electromagnetic current ($F_1(0)=1)$,
$F_A=-F_4/q^2$ is the so-called anapole moment, while $F_2$ and
$F_3$ parametrize the usual  magnetic and electric dipole moments, respectively, i.e.
\begin{equation}
F_2(0)=a_f,\quad d_f=\frac{e}{2m_f}F_3(0).
\label{F2F3}
\end{equation}

The  electric dipole moment is a  CP violating magnitude, whose value
vanishes in the SM  up to three loops for leptons. Observables 
able to disentangle it from the rest have been studied in
~\cite{Bernabeu:2004ww,nos07} using techniques  similar to  those
presented in the present paper; therefore, it is not  going to be
considered here.  The P-odd anapole  moment differs from zero due
to weak corrections.   In cross-sections its contribution  will be suppressed
by factors of $q^2/M_Z^2$ compared  to the leading QED corrections, so
that  its  determination will  remain  below  the  sensitivity of  the
proposed observables.
As emphasized already in the Introduction,
and as is clear from Eq.~(\ref{F2F3}),
$F_2(q^2)$ coincides with the anomaly $ a_\tau$ only at $q^2=0$.
In Super  B factories the squared center-of-mass energy
$s = q^2 \approx (10\, GeV)^2$, and  therefore
$F_2(q^2)$ is no longer the magnetic anomaly.

When attempting to extract the value of $F_2$ from scattering experiments
(as opposed to using, say, a background magnetic field) one encounters
additional complications due to the contributions of various other
Feynman graphs, not related to the magnetic form factor.
For example, in the case of
$e^+e^-\longrightarrow \tau^+\tau^-$ that we will consider,
one receives contributions not only from the usual  $s$-channel one-loop vertex corrections
but also from box diagrams. The contributions of the latter may interfere
in the experimental determination of what we call $F_2(q^2)$,
i.e. the magnetic part coming {\it only} from the vertex, and should be
somehow ``subtracted out''. This may be done either by computing the box contributions
and subtracting them from the cross-section, or by performing the
measurement in a kinematic region where the boxes happen to be numerically subleading.
The strategy we propose in this paper for eliminating the contamination from the boxes
is to measure the observables on top of the $\Upsilon$ resonances; in this kinematic regime
the (non-resonant) box diagrams are numerically negligible,
and only one loop corrections to the $\gamma f\bar{f}$ vertex are relevant.

The
direct text-book
computation of the magnetic part of the standard one-loop QED vertex yields ~\cite{Itzykson:1980rh}
\bea
&&F_2(s)=\left(\frac{\alpha}{2\pi}\right)\frac{2m_\tau^2}{s}\frac{1}{\beta}\left(\log\frac{1+\beta}{1-\beta}-i\,\pi\right),\quad\mbox{for}
\quad q^2 = s > 4 m_\tau^2,\label{MMff}
\eea
where $\alpha$ is the fine structure constant and $\beta = \left(1-4m_\tau^2 / s\right)^{1/2}$ is the
velocity of the $\tau$.
For $M_\Upsilon \sim 10\, {\rm GeV}$,
\beq
F_2(M_\Upsilon^2)=(2.65 - 2.45\, i)\times 10^{-4}.\label{f2}
\eeq
Evidently,  at this energy the real and imaginary parts are of the same order.

Note that the above expression for $F_2(s)$ is gauge-independent, despite being an off-shell amplitude.
This fact may be easily verified through an explicit calculation of the vertex diagram;
more generally, the gauge-independence of $F_2(s)$ may
be understood in terms of the way the gauge-cancellations organize themselves
in the QED $S$-matrix elements. Specifically, the vacuum polarization is gauge-independent by itself;
the gauge-dependence of the direct box cancels exactly against that of the
crossed box; the gauge-dependence of the vertex correction can therefore cancel only
against the fermion self-energy graphs renormalizing the external (on-shell) fermions.
The latter however are proportional
to $\gamma_{\mu}$. Therefore the contribution of the vertex proportional to
$\sigma_{\mu\nu}q^{\nu}$ must be individually gauge-independent~\cite{foot2}.

\section{$e^+e^-\longrightarrow \tau^+\tau^-$ at Super B Factories.}\label{section:gamma}
\begin{figure}[hbtp]
\begin{center}
\epsfig{file=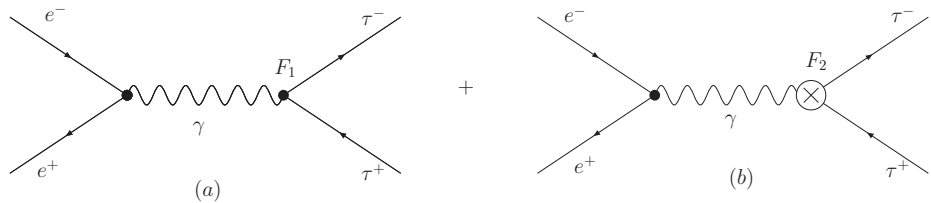,width=.9\textwidth}
\end{center}
\caption{Diagrams: (a) direct $\gamma$ exchange, (b) $F_2$ in $\gamma$ exchange. }
\label{fig:figura1}
\end{figure}

In  this section  we first  consider the  $\tau$-pair production  in 
$e^+e^-$ collisions
through   direct  $\gamma$   exchange   (diagrams  (a)   and  (b)   in
Fig. \ref{fig:figura1}). Next, we will  show that the basic results of
this section still hold for resonant $\Upsilon$ production.

The spin-independent differential cross-section for $\tau$ pair-production
can be written as:
\begin{equation}
\frac{d \sigma^0}{d  cos\theta_{\tau^-}}= \frac{\pi \alpha^2}{8 s}\beta\left[(2-\beta^2\sin^2\theta_{\tau^-})|F_1(s)|^2+4\re{F_1(s)F_2(s)^*}\right]
\label{cross01}
\end{equation}
Note that, at one loop, we have the identity
\beq
\re{F_1(s)F_2(s)^*}=\re{F_2(s)}\,.
\eeq
The $\theta_{\tau^-}$ angle is determined  in the center-of-mass (CM) frame
by the  outgoing $\tau^-$  and the incoming  $e^-$ momenta. As  can be
seen  from   Eq.~(\ref{cross01}),   a  precise  measurement   of  the
$\theta_{\tau^-}$ angle  allows to fit the  differential cross-section
and obtain a measurement for the $\re{F_2(s)}$ form  factor. In B/Flavor factories, this procedure
requires that the  $\tau$ production plane and direction  of flight are
fully reconstructed. In Ref.~\cite{kuhn}  it is shown that this can be
achieved if  both $\tau$'s decay semileptonically.
Following these ideas, the differential cross-section for $\displaystyle e^+e^-\longrightarrow \gamma \longrightarrow \tau^+(\vec{s}_+)\; \tau^-(\vec{s}_-)
\longrightarrow h^+\bar{\nu}_\tau\,  h^-\nu_\tau$ can then be written as~\cite{tsai}:
\begin{eqnarray}
\frac{d \sigma}{d  cos\theta_{\tau^-}}
&=& \frac{\pi\; \alpha^2}{2s}\beta\:  \left[(2-\beta^2\sin^2\theta_{\tau^-})\; |F_1(s)|^2+4\, \, \re{F_2(s)}\right]
\nonumber \\
&\times& Br(\tau^- \rightarrow h^-\nu_\tau)\;Br(\tau^+ \rightarrow h^+\bar{\nu}_\tau)
\label{eq:cros01}
\end{eqnarray}
The real part of $F_2$ can be measured by a fit to Eq.~(\ref{eq:cros01}),
with a sensitivity that will be given basically by the
statistical error, smeared by the precision in the determination of the angle of the outgoing $\tau$.
Results presented in Table 1  
of Section 6 only consider statistical errors. The integrated cross-section from Eq. (14) will provide, to leading order in $\alpha$ ($F_1=1$,   
$F_2=0$), the normalization for all the asymmetries considered in this paper.

As can be seen from Eq. (\ref{eq:cros01}), the differential cross-section
does not depend on the imaginary part of $F_2$, as expected.
The imaginary
part of $F_2$ is a T-odd, C- and P-even quantity; therefore, a suitable
observable to look for its effects will be the normal (to the scattering
plane) polarization of the outgoing $\tau$, as we will show in what follows.

\subsection{Normal Asymmetry}\label{section:na}

 In order to have sensitivity
to the $\tau$ polarization, one has to measure the angular distribution of the decaying particles.
In fact, in the cross-section all information on the imaginary part of $F_2$ is carried out by the
linear spin-terms:
\begin{equation}
\frac{d \sigma^{S}}{d
cos_{\tau^-}}=\frac{\pi \alpha^2}{4\ s}\; \beta
\; (s_-+s_+)_y\: Y_+\label{cross2}
\end{equation}
where
\beq
Y_+=\gamma\: \beta^2\left(\cos \theta_{\tau^-}\, \sin\theta_{\tau^-}\right)\, \im{F_2(s)}, \label{cross3}
\eeq
and $\gamma = \sqrt{s} /2 m_\tau$ is the dilation factor.

We work in the CM frame of reference, and the orientation of the coordinate system is
the same as in Ref.\cite{Bernabeu:2004ww}.
The $\mbox{\boldmath $s$}_\pm$  vectors are the
 $\tau^\pm$  spin vectors in the $\tau^\pm$ rest system, $s_\pm=(0, s^x_\pm, s^y_\pm, s^z_\pm)$. Polarization along the directions $x,y,z$ correspond
 to what is called transverse, normal, and longitudinal polarizations, respectively.
Eq.(\ref{cross3}) shows that the contribution of
the chirality flipping $F_2$ to the normal polarization
is enhanced by a factor of $\gamma$ with respect to
other possible non-chirality flipping contributions.

The polarization of the final fermion ($\tau^\pm$) can be studied
by looking at the angular distribution of its decay products.
This again requires that only
semileptonic $\tau$ decays must be considered~\cite{kuhn}. The cross-section is \cite{tsai}:
\begin{eqnarray}
&&d\sigma \left(e^+e^-\rightarrow \gamma
\rightarrow \tau^+(\vec{s}_+)\tau^-(\vec{s}_-) \rightarrow h^+\bar{\nu}_\tau h^-\nu_\tau\right)=
\frac{d\Omega_{h^+}}{4\pi}\, \frac{d\Omega_{h^-}}{4\pi}\nonumber\\
&&\times\: 4\, d\sigma
\left(e^+e^- \rightarrow \tau^+(\vec{n}_+^*)
\, \tau^-(\vec{n}_-^*)\right)
\, Br(\tau^+ \rightarrow h^+\bar{\nu}_\tau)
Br(\tau^- \rightarrow h^-\nu_\tau)\,,
\label{eq:cros1}\end{eqnarray}
with
\beq \overrightarrow{n}_\pm^*= \mp\alpha_\pm
\frac{\overrightarrow{q}^{  *}_\pm}{
\arrowvert\overrightarrow{q}^{  *}_\pm\arrowvert} =
\mp\alpha_\pm(\sin\theta_{\pm}^*\, \cos\phi_\pm,
\sin\theta_{\pm}^*\, \sin\phi_\pm,\cos\theta_{\pm}^*)\,.\nonumber\\
\eeq

The $\phi_{\pm}$ and $\theta^*_{\pm}$ angles are the azimuthal and polar angles of the
produced hadrons $h^\pm$ ($\hat{q}^*_{\pm}$) in the $\tau^\pm$
rest frame (the * means that the quantity is
given in the $\tau$ rest frame) and $\alpha_{h}$ is the
 polarization analyzer.

To preserve the normal polarization term in the cross-section of Eq.~(\ref{eq:cros1}) one has to define a particular
integration over the angular variables defined before. Indeed, the usual integration over the complete range of the $\tau^-$ variables
$d\Omega_{\tau^-}$ erases all the information on the $Y_+$ term in the cross-section, so we must perform an asymmetric forward-backward (FB)
integration on the $\theta_\tau$ angle. This can be done by defining
\bea
\sigma_{FB}(\vec{s}_+,\vec{s}_-)&\equiv& 2\pi\left\{\int_0^1 d\left(\cos\theta_{\tau^-}\right)\left[\frac{d \sigma}{d
\Omega_{\tau^-}}\; \right]-\int_{-1}^0 d\left(\cos\theta_{\tau^-}\right)\left[\frac{d \sigma}{d
\Omega_{\tau^-}}\; \right]\right\}\nonumber\\
&=&\frac{\pi\;\alpha^2}{6\ s}\; \beta^3\; \gamma\; (s_-+s_+)_y\: \im{F_2(s)}\label{anormal}\,,
\eea
which retains the $\im{F_2}$ term.
Then, the cross-section in Eq.~(\ref{eq:cros1}) can be written as
\begin{eqnarray}
d^4\sigma_{FB} &=&
\frac{2\pi\alpha^2}{3\, s}
\, Br(\tau^+ \rightarrow h^+\bar{\nu}_\tau)
Br(\tau^- \rightarrow h^-\nu_\tau) \, \frac{d\Omega_{h^+}}{4\pi}\,
\frac{d\Omega_{h^-}}{4\pi}\, \nonumber\\
&&\times \, \left[(n_-^*)_y+(n_+^*)_y\right]\: \beta^3\: \gamma\, \im{F_2(s)}\,. \label{eqxy1}
\end{eqnarray}

Integrating over as many
kinematic variables as possible ($\theta^*_{\pm}$), without erasing the information on $F_2$,
we finally find that the
differential cross-section can be written as
\bea
\frac{d\sigma_{FB}}{d\phi_\pm}
&=& \mp  \frac{\pi\; \alpha^2}{12 s}\,
Br(\tau^+ \rightarrow h^+\bar{\nu}_\tau)
Br(\tau^- \rightarrow h^-\nu_\tau)\nonumber \\
&&\times (\alpha_\pm) \beta^3\: \gamma\, \im{F_2(s)}\, \sin\phi_\pm \,.
\eea

To get an observable sensitive to the relevant signal, we must now define the
azimuthal normal asymmetry as:
\begin{equation}
A_N^{\pm} =\frac{\sigma^\pm_{L} - \sigma^\pm_{R}}{\sigma}=
\pm\, \alpha_\pm\, \frac{1}{2(3-\beta^2)}\, \beta^2\, \gamma\, \im{F_2(s)}
\label{asymN}
\end{equation}
where
\beq
\sigma^\pm_L \equiv \int_{\pi}^{2\pi}d\phi_\pm \, \left[
\frac{d\sigma_{FB}}{d\phi_\pm}\right],\quad
\sigma^\pm_R \equiv \int_{0}^{\pi}d\phi_\pm \, \left[
\frac{d\sigma_{FB}}{d\phi_\pm}\right]=-\sigma^\pm_L
\eeq

\section{Polarized $e^-$ beams.}

For polarized electrons, available at Super B factories,
an alternative procedure for measuring $F_2$
would be to consider the longitudinal and transverse polarizations of the
outgoing $\tau$'s.

Because $F_1$ and $\re{F_2}$ have the same properties under C, P and T symmetries,
any single observable sensitive to one will also carry information on the other.
Therefore, the extraction of the chirality flipping $\re{F_2}$ requires
two independent observables, where $F_1$ and $F_2$ enter with different
coefficients; that would allow to express $F_2$
as a linear combination of the two measured observables. Following the notation and
procedure of references~\cite{arcadi,nos07,nos}, our aim is to build observables that are linear in $F_2$.

$\re{F_2}$ is even  under T, C and P, while  the longitudinal and transverse
(to the  scattering plane) polarizations  of each $\tau$ are  the only
components of  the spin matrix  that are even  under T and C,  but odd
under P.  For this reason,  an observable sensitive to $F_2$ will need
an   additional  P-odd   contribution  coming,   in  our   case,  from
longitudinally polarized electrons.

The linear spin-dependent part of the differential cross-section for $\tau$ pair production, with
polarized electrons with helicity $\lambda$, can be written as
\bea
\left.\frac{d \sigma^{S}}{d
\cos_{\tau^-}}\right|_\lambda&=&\frac{\pi \alpha^2}{8\ s}\, \beta\, \bigg\{\, (s_-+s_+)_y\: Y_+\nonumber\\
&&\qquad\qquad +\lambda\left[
(s_-+s_+)_x\: X_++(s_-+s_+)_z\: Z_+\right]\, \bigg\}\,, \label{crossp2}
\eea
with
\bea
X_+&=&\sin\theta_{\tau^-}\left[|F_1|^2+(2-\beta^2)\gamma^2\; \re{F_2}\right]\frac{1}{\gamma}\,, \nonumber \\
Z_+&=&\cos \theta_{\tau^-}\left[|F_1|^2+2\; \re{F_2}\right]\,, \label{crossp3}
\eea
and $Y_+$ as defined in Eq.~(\ref{cross3}). Notice: i) The 
combination of the two form factors is different for the transverse and 
longitudinal $\tau$ polarization terms; ii) these two terms have different 
angular dependence.

As can be seen from Eq.(\ref{crossp3}),  $\re{F_2}$ contributes linearly to the longitudinal and transverse
$\tau$ polarization. It may again be
observed that, due to the fact that $F_2$ is a chirality-flip form factor,
its contribution to the transverse polarization is enhanced by the factor
$\gamma^2$ with  respect to the chirality-non flipping factor $F_1$.
This a very important fact for our purposes,
because it will allow the accurate determination of $F_2$.

\subsection{Transverse Asymmetry}

Following a procedure similar to the one presented in section \ref{section:na}, it can be
seen    that   the   integration    over   the    $\tau^-$   variables
$d\Omega_{\tau^-}$ erases from Eq.~(\ref{crossp3})
all information on the $Z_+$ and $Y_+$ terms. Then, the  differential cross-section  for the
process
$\displaystyle e^+e^-|_{{\rm Pol}}\rightarrow
\gamma \rightarrow \tau^+(\vec{s}_+)\; \tau^-(\vec{s}_-) \rightarrow h^+\bar{\nu}_\tau h^-\nu_\tau$
is given by
\begin{eqnarray}
\left.d^4\sigma^{S}\right|_\lambda &=&
\frac{\pi^2\alpha^2\beta}{2\, s}
\, Br(\tau^+ \rightarrow h^+\bar{\nu}_\tau)
Br(\tau^- \rightarrow h^-\nu_\tau) \, \frac{d\Omega_{h^+}}{4\pi}\,
\frac{d\Omega_{h^-}}{4\pi}\, \nonumber\\
&&\times \beta \lambda \left[(n_-^*)_x+(n_+^*)_x\right]
\frac{1}{\gamma}\left[|F_1|^2+(2-\beta^2)\gamma^2 \re{F_2}\right] \label{eqxyp1}\,.
\end{eqnarray}

Subtracting the cross-sections for different helicities,\footnote{This subtraction eliminates also
higher order absorptive parts that may be present.
See Ref. \cite{nos07}}
\beq
\left.\d^2\sigma^{S}\right|_{{\rm Pol}( e^-)}\equiv \frac{1}{2}\left[\, \left.d^4\sigma\right|_{\lambda=1}-
\left.d^4\sigma\right|_{\lambda=-1}\, \right]\,,
\label{spol1}
\eeq
and integrating over as many
kinematic variables as possible, without erasing the information on the $F_2$ term, we get
\bea
\left.\frac{d\sigma^{S}}{d\phi_\pm}\right|_{{\rm Pol}( e^-)}
&=& \mp  \frac{\pi^2\alpha^2\beta}{16 s}
Br(\tau^+ \rightarrow h^+\bar{\nu}_\tau)
Br(\tau^- \rightarrow h^-\nu_\tau)\nonumber \\
&&\times \frac{1}{\gamma}\left[|F_1|^2+(2-\beta^2)\gamma^2
\re{F_2}\right] \left[(\alpha_\pm) \cos\phi_\pm\right]\,.
\eea

To get an observable sensitive to the relevant signal define the
azimuthal transverse asymmetry as
\bea
A_T^{\pm} &=&\frac{\sigma^\pm_{R}|_{{\rm Pol}} - \sigma^\pm_{L}|_{{\rm Pol}}}{\sigma}
\nonumber \\
&=&\mp\, \alpha_\pm\, \frac{3\pi}{8(3-\beta^2)\gamma}\left[|F_1|^2+(2-\beta^2)\gamma^2 \re{F_2}\right]\,,
\label{asymt}
\eea
where
\begin{eqnarray}
\sigma^\pm_L|_{{\rm Pol}} &\equiv& \int_{\pi/2}^{3\pi/2}d\phi_\pm \, \left[\left.
\frac{d\sigma^S}{d\phi_\pm}\right|_{{\rm Pol}(e^-)}\right]=\; \pm \; Br(\tau^+ \rightarrow h^+\bar{\nu}_\tau)
Br(\tau^- \rightarrow h^-\nu_\tau) \nonumber\\
&&\quad\times \; \alpha_\pm\frac{(\pi\alpha)^2\beta}{8s}\, \frac{1}{\gamma}\:
\left[|F_1|^2+(2-\beta^2)\gamma^2 \re{F_2}\right]\,, \label{sigmaL}\\
\sigma^\pm_R |_{{\rm Pol}}&\equiv& \int_{-\pi/2}^{\pi/2}d\phi_\pm \, \left[\left.
\frac{d\sigma^S}{d\phi_\pm}\right|_{{\rm Pol}(e^-)}\right]=-\sigma^\pm_L|_{{\rm Pol}}\,.\label{sigmaR}
\end{eqnarray}

It is clear from Eq.~(\ref{asymt}) that in order 
to separate out $\re{F_2}$, we need to remove the  contribution of $F_1$. This can be
done in  two ways. The first is  to use that $|F_1|^2=1$ at tree-level,
and assume that any additional contribution to $F_1$
will be of the same order as $F_2$;
since $F_2$ is a chirality flipping quantity, it will be
enhanced  by a factor $\gamma^2$ with respect to the additional $F_1$.
Under these assumptions, a measurement of the
$A_T$ asymmetry  (subtracted with the tree level value $|F_1|^2 = 1$) 
will translate into a $\re{F_2}$ measurement. 
The second way  is  to consider a  new observable,  relating
$|F_1|^2$ and $\re{F_2}$,  and combine the two measurements  to extract the
value  of $\re{F_2}$.  This can  be  done by  defining a  longitudinal
asymmetry ($A_L$) as follows.

\subsection{Longitudinal Asymmetry}

From Eqs.~(\ref{crossp2}) and (\ref{crossp3}) it can be seen that,
as happened with Eq.~(\ref{anormal}) for the normal asymmetry,  an asymmetrical (FB)
integration  on the $\theta_\tau$ angle will select the longitudinal term of the cross-section
\beq
\hspace{-0.5cm}
\left. \sigma_{FB}^S(\vec{s}_+,\vec{s}_-)\right|_\lambda\equiv
\frac{\pi\alpha^2}{4\ s}\; \beta\left[\lambda\;  (s_-+s_+)_z\widetilde{Z}_++
\frac{2}{3}\beta^2\gamma(s_-+s_+)_y\; \im{F_2}\right]\,,
\label{crossp4}
\eeq
where $\widetilde{Z}_+ = Z_+ / \cos \theta_{\tau^-}$.
Following a similar procedure as in the previous paragraph,
and after subtracting for different helicities as was done in Eq. (\ref{spol1}),
but integrating over the azimuthal
angles $\phi_\pm$ instead of the polar $\theta^*_\pm$ ones,
we end up with the final expression for the asymmetrical
(FB)  differential cross-section for polarized electrons:

\bea
\left.\frac{d\sigma^{S}_{FB}}{d(\cos\theta_\pm^*)}\right|_{{\rm Pol}( e^-)}
&=& \mp  \frac{\pi\alpha^2\beta}{2 s}
Br(\tau^+ \rightarrow h^+\bar{\nu}_\tau)
Br(\tau^- \rightarrow h^-\nu_\tau)\nonumber \\
&&\times \left[|F_1|^2+2\; \re{F_2}\right] \left[(\alpha_\pm) \cos\theta^*_\pm\right]\,.
\eea
Then, we define the longitudinal asymmetry as

\bea
A_L^{\pm} &=&\frac{\sigma^\pm_{FB}(+)|_{{\rm Pol}} - \sigma^\pm_{FB}(-)|_{{\rm Pol}}}{\sigma}\nonumber \\
&&\qquad =\mp\, \alpha_\pm\, \frac{3}{4(3-\beta^2)}\left[|F_1|^2+2\; \re{F_2}\right]\,,
\label{asyml}
\eea
where
\begin{eqnarray}
\sigma^\pm_{FB}(+)|_{{\rm Pol}}&\equiv&\int_{0}^{1}d(\cos\theta_\pm^*) \, \left.
\frac{d\sigma^S_{FB}}{d(\cos\theta_\pm^*)}\right|_{{\rm Pol}(e^-)}=\mp \alpha_\pm\,
Br(\tau^+ \rightarrow h^+\bar{\nu}_\tau)\nonumber\\
&\times&
Br(\tau^- \rightarrow h^-\nu_\tau)\; \frac{\pi \alpha^2}{4s}\beta\,
\left[|F_1|^2+2\; \re{F_2}\right] \label{sfbp}\\
\sigma^\pm_{FB}(-)|_{{\rm Pol}} &\equiv& \int_{-1}^{0}d(\cos\theta_\pm^*)\, \left.
\frac{d\sigma^S_{FB}}{d(\cos\theta_\pm^*)}\right|_{Pol (e^-)}=-\sigma^\pm_{FB}(+)|_{{\rm Pol}}\,.
\label{sfbm}
\end{eqnarray}

Combining Eq.(\ref{asymt}) and Eq. (\ref{asyml}) one can determine the real part of $F_2(s)$. Specifically,
\beq
\re{F_2(s)}=\mp\frac{8(3-\beta^2)}{3\pi\gamma\beta^2}\frac{1}{\alpha_\pm}
\left(A_T^\pm-\frac{\pi}{2\gamma}A_L^\pm\right)\,.\label{combined}
\eeq

\section{Observables on the $\Upsilon$ resonance}

\begin{figure}[hbtp]
\begin{center}
\epsfig{file=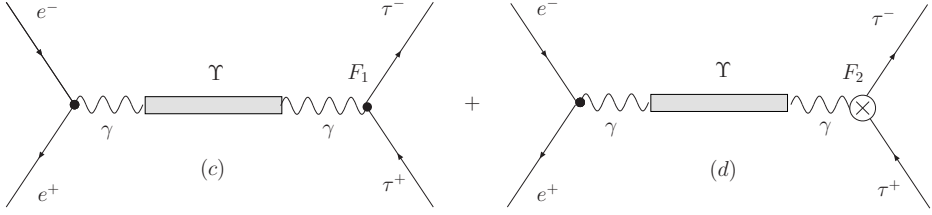,width=.9\textwidth}
\end{center}
\caption{Diagrams: (c) $\Upsilon$ production, (d) $F_2$ in $\Upsilon$ production}
\label{fig:fig3}
\end{figure}

As explained in the Introduction, our aim is to measure the observables on the top of the $\Upsilon$ peak
where the $\tau$ pair-production is  mediated by the resonance.
The leading diagrams for the process $e^+e^- \rightarrow \Upsilon \rightarrow
\tau^+\tau^-$ are shown in Fig.\ref{fig:fig3}. Given that we are interested in $\tau$ pairs produced by the
 decays of the $\Upsilon$ resonances, we can use $\Upsilon(1S)$ and $\Upsilon(2S)$, since
their decay rates into $\tau$ pairs have been measured and are sizeable. We assume
that only the resonant diagrams (c) and (d) of Fig. \ref{fig:fig3}
dominate the process on the $\Upsilon $ peaks, so no contribution
from box diagrams has to be considered.
%
%
As discussed in ref.~\cite{Bernabeu:2004ww}, the $\tau$ pair-production at the $\Upsilon$
peak introduces the same $\tau$ polarization matrix terms as the
direct production with a $\gamma$ exchange
(diagrams (a) and (b), Fig. \ref{fig:figura1}) that we have calculated in section \ref{section:gamma}.
The only difference is an overall factor $|H(s)|^2$
which is responsible for the enhancement of the cross-section  at resonant
energies; the pure resonant  amplitude is given by
\beq H(M_\Upsilon^2) =\frac{4\pi\alpha Q^2_b}{M_\Upsilon^2}\frac{\left|F_\Upsilon\left(M_\Upsilon^2\right)\right|^2}{i\Gamma_\Upsilon M_\Upsilon}=-i \,
\frac{3}{\alpha}
Br\left(\Upsilon \rightarrow e^+e^-\right)\,.
\label{factor}
\eeq
At the $\Upsilon$ peak,
the interference of diagrams (a) and (d),
plus the interference of diagrams (b) and (c), shown in Fig. \ref{fig:figura1} and \ref{fig:fig3}, is
exactly zero, and so is the interference of diagrams (a) and (c).
Finally, the only contributions proportional to the $F_2$
come  from the interference of diagrams (c) and (d), while diagram
(c) squared gives the leading contribution to the cross-section.

\subsection{$Z$ contribution on the $\Upsilon$ resonance.}

\begin{figure}[hbtp]
\begin{center}
\epsfig{file=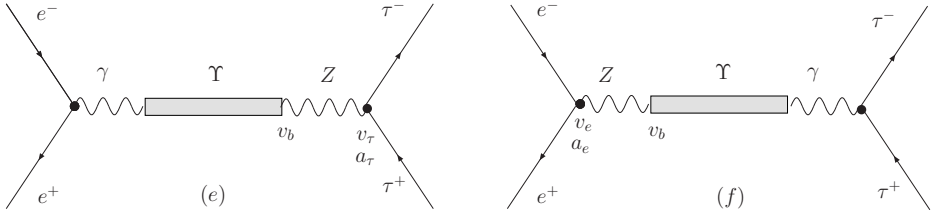,width=.9\textwidth}
\end{center}
\caption{$\gamma$ and $Z$ interchange on $\Upsilon$ production}
\label{fig:fig4}
\end{figure}

It is important to notice that the observables previously defined
receive contributions also from the standard $Z-\gamma$ interference, computed at the resonance.
The process has been studied in detail in ref.~\cite{bernabeu.pascual};
 the dominant (resonant) contribution will come from
the interference of diagram (c) in Fig.\ref{fig:fig3} with diagrams (e) and (f) in
 Fig.\ref{fig:fig4}.
%
\begin{figure}[hbtp]
\begin{center}
\epsfig{file=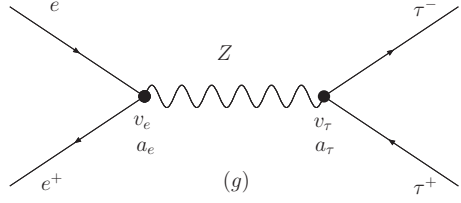,width=.45\textwidth}
\end{center}
\caption{Non-resonant $Z$ interchange}
\label{fig:fig5}
\end{figure}

Following a procedure similar to that
explained in detail in \cite{Bernabeu:2004ww},
one can find the following relations among the amplitudes of diagrams (c), (e) and
(f) with the non-resonant  $\gamma$-mediated diagram (a) and $Z$-mediated diagram (g) of Fig. \ref{fig:fig5}.
\bea
T_c&=&T_a\times H(M_\Upsilon^2)\,,\quad T_d=T_b\times H(M_\Upsilon^2)\,,\nonumber \\
T_e&=&T_g(a_e\rightarrow 0,v_e\rightarrow v_b)\times \left(H(M_\Upsilon^2)\frac{Q_e}{Q_b}\right)\,, \nonumber \\
T_f&=&T_g(a_\tau\rightarrow 0,v_\tau\rightarrow v_b)\times\left( H(M_\Upsilon^2) \frac{Q_e}{Q_b}\right)\,.
\eea

From these relations, the resonant $\gamma-\Upsilon-Z$ interference
can be extracted from the purely $\gamma-Z$ interference. This was studied
in Ref.\cite{nos07} and the additional contributions to the spin-averaged cross-section
and to its spin- dependent part are
\bea
\left.\frac{d \sigma^0}{d  \Omega_{\tau^-}}\right|_\lambda^{Z}
&=&-\frac{\alpha^2 \beta}{8\ (2s_wc_w)^2}\left|H\left(M_\Upsilon^2\right)\right|^2\frac{Q_e}{Q_b} v_b\; |P_Z(M_\Upsilon^2)|^2
\nonumber\\
&&\qquad\qquad\qquad\qquad\qquad\qquad\qquad\times\left(M_\Upsilon^2-M_Z^2\right)\;  M^{Z}_0 \,,\label{z0} \\
\left.\frac{d \sigma^S}{d  \Omega_{\tau^-}}\right|_\lambda^{Z}
&=&\frac{\alpha^2\beta}{8\ (2s_wc_w)^2}\left|H\left(M_\Upsilon^2\right)\right|^2\frac{Q_e}{Q_b} v_b\; |P_Z(M_\Upsilon^2)|^2,\nonumber \\
&\times&
\left\{\Gamma_Z\, M_Z \:(s_-+s_+)_y \: Y_+^{Z}\right.\nonumber\\
&+&\left.\left(M_\Upsilon^2-M_Z^2\right)\left[(s_-+s_+)_x\: X^{Z}_+
+(s_-+s_+)_z\: Z^{Z}_+\right]\right\}, \label{zs}
\eea
where
\bea
&&|P_Z(M_\Upsilon^2)|^2
=\left[(M_\Upsilon^2-M_Z^2)^2+\Gamma_Z^2M_Z^2\right]^{-1},\nonumber \\
&&a=-\frac{1}{2},\quad
v=-\frac{1}{2}+2s_w^2\quad v_b=-\frac{1}{2}+\frac{2}{3}s_w^2,\quad Q_b=\frac{-1}{3},\quad Q_e=-1,\nonumber \\
&&M_0^{Z}=\lambda\, a(2\beta\cos\theta_{\tau^-}+2-\beta^2\sin^2\theta_{\tau^-})-2v(2-\beta^2\sin^2\theta_{\tau^-}),\nonumber
\\
&&X_+^{Z}=\frac{1}{\gamma}\left[4\lambda v-a\,(2+\beta\cos\theta_{\tau^-})\right]\sin\theta_{\tau^-},\nonumber \\
&&Z^{Z}_+=4\lambda v\cos\theta_{\tau^-}-a\,
\left[\beta(1+\cos^2\theta_{\tau^-})+2\cos\theta_{\tau^-}\right],\nonumber \\
&&Y_+^{Z}= -\frac{\lambda}{\gamma}\beta\, a \sin\theta_{\tau^-}.
\label{crosz}
\eea

To obtain the contribution of the $Z$-interference to both the cross-section and
the normal asymmetry for unpolarized electrons, one must average over
$\lambda$ helicities in Eqs.(\ref{z0}) and (\ref{zs}). Then,
\bea
\left.\frac{d \sigma^0}{d  \Omega_{\tau^-}}\right|^Z
&=&\frac{\alpha^2\beta}{16\ M_\Upsilon^2}\left|H\left(M_\Upsilon^2\right)\right|^2\; (2-\beta^2\sin^2\theta_{\tau^-})\nonumber \\
&&\qquad\times\underbrace{\frac{4v}{(2s_wc_w)^2}\frac{Q_e\: v_b}{Q_b}\: M_\Upsilon^2\:
\left(M_\Upsilon^2-M_Z^2\right)\, |P_Z(M_\Upsilon^2)|^2}_\zeta,\label{zcross1}\\[.5\baselineskip]
\left.\frac{d \sigma^S}{d  \Omega_{\tau^-}}\right|^Z
&=&-\frac{\alpha^2\beta\; a}{8\ (2s_wc_w)^2}\left(\left|H\left(M_\Upsilon^2\right)\right|^2\frac{Q_e}{Q_b} v_b\right)\; |P_Z(M_\Upsilon^2)|^2\;
\left(M_\Upsilon^2-M_Z^2\right)\nonumber\\
&&\times\bigg\{(s_-+s_+)_x\frac{1}{\gamma}(2+\beta\cos\theta_{\tau^-})\nonumber\\
&&+(s_-+s_+)_z\left[\beta(1+\cos^2\theta_{\tau^-})+2\cos\theta_{\tau^-}\right]\bigg\}.\label{zcross2}
\eea

Eq.~(\ref{zcross1})  shows that the $Z$
contribution will only enter in  the determination of $F_1$, with the suppression
factor $\zeta\approx -1.576\times 10^{-3}$. Furthermore, it is controlled
by the small vector neutral coupling $v$ to leptons. This implies that the angular distribution is 
like that for $|F_1|^2$ and the measurement of $\re{F_2}$ from the cross-section is not modified.

Similarly, Eq. (\ref{zcross2}) shows that there is no contribution from the $Z$-interference  to the
normal asymmetry, and the extraction of the $\im{F_2}$ value from this observable through 
Eq. (\ref{asymN}) is again not modified.

For polarized electrons, subtracting the cross-sections for different helicities and
integrating, as was done in the previous section for the transverse asymmetry, one obtains
the new contribution from Z-interference to Eqs. (\ref{sigmaL}) and (\ref{sigmaR}) as
\bea
\left.^{Z}\sigma^\pm_L\right|_{{\rm Pol}} &=& \frac{2\pi \alpha^2\beta}{(2s_wc_w)^2}\, |P_Z(M_\Upsilon^2)|^2 \;
Br(\tau^+ \rightarrow h^+\bar{\nu}_\tau) Br(\tau^-
\rightarrow h^-\nu_\tau)\nonumber \\
&\times&\left(M_\Upsilon^2-M_Z^2\right)\left|H\left(M_\Upsilon^2\right)\right|^2\frac{Q_e}{Q_b} v_b\;
\left[\left(\frac{\beta^2}{3}-1\right)a\; \pm \frac{\pi v}{4\gamma}\: \alpha_\pm
\right]\,,\\
\left. ^{Z}\sigma^\pm_R\right|_{{\rm Pol}} &=& \frac{2\pi \alpha^2\beta}{(2s_wc_w)^2 }\, |P_Z(M_\Upsilon^2)|^2\;
Br(\tau^+ \rightarrow h^+\bar{\nu}_\tau) Br(\tau^- \rightarrow h^-\nu_\tau)\nonumber \\
&\times&\left(M_\Upsilon^2-M_Z^2\right)\left|H\left(M_\Upsilon^2\right)\right|^2\frac{Q_e}{Q_b} v_b\;
\left[\left(\frac{\beta^2}{3}-1\right)a\; \mp \frac{\pi v}{4\gamma}\: \alpha_\pm\right]\,,
\eea
so that the azimuthal transverse asymmetry, $A_T^\pm$ of Eq. (\ref{asymt}), reads
\beq
^{Z}A_T^{\pm} =\pm \alpha_\pm\frac{3\pi(2-\beta^2)\gamma}{8(3-\beta^2)}\,
\overbrace{\left(\frac{4 v_b\: v\; M_\Upsilon^2(M_\Upsilon^2-M_Z^2)}{\gamma^2(2s_wc_w)^2(2-\beta^2)Q_b}
|P(M_\Upsilon^2)|^2 \right)}^\varepsilon \,,
\label{asymzt}
\eeq

Similarly, the $Z$ contribution to the asymmetric cross-sections of (\ref{sfbp}) and (\ref{sfbm})
for the longitudinal polarization of $\tau$'s with polarized electrons is
\bea
\left. ^{Z}\sigma^\pm_{FB} (+)\right|_{{\rm Pol}} &=& -\frac{\pi \alpha^2\beta}{(2s_wc_w)^2}\, |P_Z(M_\Upsilon^2)|^2 \;
Br(\tau^+ \rightarrow h^+\bar{\nu}_\tau) Br(\tau^-
\rightarrow h^-\nu_\tau)\nonumber \\
&\times&\left(M_\Upsilon^2-M_Z^2\right)\left|H\left(M_\Upsilon^2\right)\right|^2\frac{Q}{Q_b} v_b\;
\left[\beta\; a\; \pm v\: \alpha_\pm \right]\\
\left. ^{Z}\sigma^\pm_{FB} (-)\right|_{{\rm Pol}} &=& -\frac{\pi \alpha^2\beta}{(2s_wc_w)^2 }\, |P_Z(M_\Upsilon^2)|^2\;
Br(\tau^+ \rightarrow h^+\bar{\nu}_\tau) Br(\tau^- \rightarrow h^-\nu_\tau)\nonumber \\
&\times&\left(M_\Upsilon^2-M_Z^2\right)\left|H\left(M_\Upsilon^2\right)\right|^2\frac{Q}{Q_b} v_b\;
\left[\beta\; a\; \mp v\: \alpha_\pm\right]\,,
\eea
and the corresponding contribution to the longitudinal asymmetry, $A_L^\pm$ of Eq. (\ref{asyml}), is given by
\beq
^{Z}A_L^{\pm} =\pm \alpha_\pm\frac{3}{4(3-\beta^2)}\,
\overbrace{\left(\frac{ 4v_b\: v\; M_\Upsilon^2(M_\Upsilon^2-M_Z^2)}{(2s_wc_w)^2Q_b}
|P(M_\Upsilon^2)|^2\right) }^{\varepsilon '}\,.
\label{asymzl}
\eeq

 At $\Upsilon$ energies, the values of the $\varepsilon$ and $\varepsilon '$ factors are of the order
 $3.17\times10^{-4}$ and $3.15\times 10^{-3}$, respectively. Contrary to what happens with the 
 non-resonant contribution
(which is two orders of magnitude smaller) this $\Upsilon$-mediated $\gamma-Z$ contribution
 must be taken into account when extracting $F_2$
from longitudinal and transverse asymmetries for polarized electrons. Nevertheless, as the 
$\gamma-\Upsilon-Z$ interference proceeds   
through the vector neutral current coupling to leptons, the structure   
of this amplitude is like the one for the contribution of the charge   
form factor $F_1$. As a consequence, the same combination (\ref{combined}) of the two   
asymmetries ($A_T-\frac{\pi}{2\gamma} A_L$) able to cancel the contribution of $|F_1|^2$   
automatically cancels the contribution of the Z interference too. We   
have thus shown that $\re{F_2}$ can be separated out from other   
contributions without any ambiguities.

\def\up{-7pt}
\def\tsize{\normalsize}
\begin{table}[hbt]{\centering
\caption{Sensitivity of the $F_2$ measurement at the $\Upsilon$ energy ($ab\, =\, {\rm atto barn}\,=\,10^{-18}b$)}
\begin{tabular}{|c|c|c|c|}
\hline
&\multicolumn{3}{|c|}{ O B S E R V A B L E}\\
\cline{2-4}
$\begin{array}{c}
\null\\[\up]
\null\\[\up]
\null\\[\up]
\null\\[\up]
\mbox{EXPERIMENT}\end{array}$
&{\tsize Cross Section}&
$\begin{array}{c}\mbox{\tsize Normal}\\[\up]\mbox{\tsize Asymmetry}\end{array}$&
$\begin{array}{c}\mbox{\tsize Transverse and}
\\[\up]\mbox{\tsize Longitudinal}\\[\up]\mbox{\tsize Asymmetry}\\[\up]\mbox{\tsize combined}^*\end{array}$\\ \cline{2-4}
$\Downarrow$&$\re{F_2}$&$\im{F_2}$&$\re{F_2}$\\ \hline
$\begin{array}{c}\mbox{\tsize Babar+Belle}\\ 2ab^{-1}\end{array}$&$4.6\times 10^{-6}$&$2.1\times 10^{-5}$&$1.0\times 10^{-5}$\\ \hline
$\begin{array}{c}\mbox{\tsize Super B/Flavor Factory}
\\[\up] \mbox{\tsize (1 yr. running)}\\ 15ab^{-1}\end{array}$&$1.7\times 10^{-6}$&$7.8\times 10^{-6}$ &$3.7\times 10^{-6}$\\ \hline
$\begin{array}{c}\mbox{\tsize Super B/Flavor Factory}
\\[\up] \mbox{\tsize(5 yrs. running) }\\ 15ab^{-1}\end{array}$&$7.5\times 10^{-7}$&$3.5\times 10^{-6}$&$1.7\times 10^{-6}$\\ \hline
\end{tabular}}

$^*$ {\footnotesize Polarized electrons required
\label{table}}
\end{table}

\section{Precision of the $F_2(M_\Upsilon^2)$ measurement and conclusions}

We can now estimate the precision that can be achieved on the determination of $F_2$
using the observables defined before. For our numerical analysis we assume a set of
integrated luminosities  for high statistics $B$/Flavor factories.
We also consider the $\pi^\pm \; \bar{\nu_\tau}$ or $\rho^\pm\;
\bar{\nu_\tau}$ ({\it i.e.} $h_1 , \;h_2= \pi ,\; \rho$ ) decay
channels for the traced $\tau^\pm$,
while we sum up over $\pi^\mp \; \nu_\tau$ and $\rho^\mp\;
  \nu_\tau$ hadronic decay channels
for the non traced $\tau^\mp$.

In Table 1 , we show the sensitivity that can be achieved for
the magnetic moment form factor  $F_2$ in different scenarios: Babar +
Belle at $2  ab^{-1}$, B/Flavor factory, 1 yr.  running ($15 ab^{-1}$)
and 5  yrs. running  ($75 ab^{-1}$).  The  results presented  in Table 1
only consider statistical  errors.  Almost all the defined
observables  show similar accuracy  in the determination of  $F_2$, but
only   the   normal   asymmetry   is   sensitive   to   its   imaginary
part. Sensitivities coming from  the cross-section require an accurate
determination of the $\theta_{\tau^-}$ angle of the outgoing $\tau$ for
a large variety of angles.

To summarize, we  have shown  that low  energy data  makes  possible a
determination of  the $\tau$ lepton  QED effects on the  measurement of
the  $\tau$  anomalous magnetic  form  factor  $F_2$  at the  $\Upsilon$
energies. A fit  to the cross-section and a  measurement of the normal
polarization  of the  outgoing $\tau$  will  determine the  real and  the
imaginary  part of $F_2$   up to  a precision  of  $10^{-6}$. 
Compared with the QED prediction of Eq. (\ref{f2}), we see that a positive  
signal appears and it can be tested to the percent level. The
$\gamma-\Upsilon-Z$ interference will  not affect the determination of
$F_2$. Polarized electron beams also open the possibility to determine
the  value of $\re{F_2}$ by  looking  at  the  transverse  and  longitudinal
polarizations of a  single $\tau$. The precision is  again of the order
$10^{-6}$. We have identified the precise combination of the transverse  
and longitudinal $\tau$ polarizations for polarized electrons which is  
able to disentangle the anomalous magnetic moment form factor from   
the contributions of both the charge form factor and, at the same time,   
the resonant $\gamma-\Upsilon-Z$ interference.

We conclude that the measurement of these sets of observables at the upcoming Super B factories
should furnish a high accuracy determination
of the rather poorly known magnetic properties of the $\tau$ lepton.

\begin{ack}
This work has been supported by CONICYT-PDT-54/94-Uruguay, by MEC
and FEDER, under the grants FPA2005-00711 and FPA2005-01678, and
by Gene\-ralitat Valenciana under the grant ACOMP07-093.
The work of J.P. is financed by the Fundaci\'on General UV. J. B. acknowledges CERN 
Theory Unit for hospitality.
J.V. also thanks Dr. J. Salt (IFIC) for helpful conversations.
\end{ack}


\begin{thebibliography}{20}
\bibitem{foot2}
This simple organization of the gauge cancellations becomes much more
intricate in the case of non-abelian contributions; nonetheless, a
gauge-independent of-shell $F_2(s)$ can still be defined, see~\cite{PT}.
\end{thebibliography}
\end{document}